\documentstyle[preprint,prl,aps]{revtex}

\def\beq{\begin{equation}}
\def\eeq{\end{equation}}
\def\beqa{\begin{eqnarray}}
\def\eeqa{\end{eqnarray}}

\def\za{\alpha}
\def\zb{\beta}
\def\lsim{\mathrel{\raise.3ex\hbox{$<$\kern-.75em\lower1ex\hbox{$\sim$}}} }
\def\gsim{\mathrel{\raise.3ex\hbox{$>$\kern-.75em\lower1ex\hbox{$\sim$}}} }

\begin{document}
\draft
\preprint{\vbox{ \hbox{IPAS-HEP-k004} \hbox{Apr 2000} \hbox{rev. Sep 2000} } } 

\title{R-parity Violating Contribution to Neutron EDM at One-loop Order}
\author{\bf Y.-Y. Keum 
$\!\!$\footnote{E-mail: keum@phys.sinica.edu.tw}
and Otto C. W. Kong
$\!\!$\footnote{E-mail: kongcw@phys.sinica.edu.tw}
}
\address{Institute of Physics, Academia Sinica, \\
Nankang, Taipei, Taiwan 11529}
\maketitle

\begin{abstract}
We present the full result for the down squark mass-squared matrix in the
complete theory of supersymmetry without R-parity where all kind of
R-parity violating terms are admitted without bias. An optimal
parametrization, the single-VEV parametrization, is used. The major
result is a new contribution to $LR$ squark mixing, involving both
bilinear and trilinear R-parity violating couplings. Among other things,
the latter leads to neutron electric dipole moment at one-loop level.
Similiar mechanism leading to electron electric dipole moment at the
same level. We present here a short report on major features of
neutron electric dipole moment from supersymmetry without R-parity
and give the interesting constraints obtained.
\end{abstract}
\pacs{PACS index:  }


\noindent{\it Introduction.}
The minimal supersymmetric standard model (MSSM) is no doubt the most
popular candidate theory for physics beyond the Standard Model (SM). The
alternative theory with a discrete symmetry called R-parity not imposed
deserves no less attention. A complete theory of supersymmetry (SUSY) without
R-parity, where all kind of R-parity violating (RPV) terms are admitted 
without bias, is generally better motivated than {\it ad hoc} versions of 
RPV theories. The large number of new parameters involved,
however, makes the theory difficult to analyze. It has been 
illustrated\cite{k} that an optimal parametrization, called the single-VEV 
parametrization, can be of great help in making the task manageable. 
Here in this letter, we use the formulation to present the full result 
for the down squark mass-squared matrix. The major
result is a new contribution to $LR$ squark mixing, involving both
bilinear and trilinear RPV couplings. The interesting 
physics implications of this new contribution are discussed. Among such
issues, we focus here on the RPV contribution to neutron electric dipole
moment (EDM) at one-loop level. 

Neutron and electron EDM's are important topics for new CP violating
physics. Within MSSM, studies on the plausible EDM contributions lead to
the so called SUSY-CP problem\cite{susycp}. In the domain of R-parity
violation, two recent papers focus on the contributions from the 
trilinear RPV terms and conclude that there is no contribution at the 
1-loop level\cite{two}. Perhaps it has not been emphasized enough in the 
two papers that they are {\it not} studying the complete theory of SUSY without
R-parity. It is interesting to see in the latter case that there is in
fact contribution at 1-loop level, as discussed below. We would like to
emphasize again that the new contribution involves both bilinear and 
trilinear (RPV) couplings. Since various other RPV scenarios studied in
the literature typically admit only one of the two types of couplings,
the contribution has not been previously identified.

The most general renormalizable superpotential for the supersymmetric
SM (without R-parity) can be written  as
\small\beqa
W \!\! &=& \!\varepsilon_{ab}\Big[ \mu_{\alpha}  \hat{H}_u^a \hat{L}_{\alpha}^b 
+ h_{ik}^u \hat{Q}_i^a   \hat{H}_{u}^b \hat{U}_k^{\scriptscriptstyle C}
+ \lambda_{\alpha jk}^{\!\prime}  \hat{L}_{\alpha}^a \hat{Q}_j^b
\hat{D}_k^{\scriptscriptstyle C} 
\nonumber \\
&+&
\frac{1}{2}\, \lambda_{\alpha \beta k}  \hat{L}_{\alpha}^a  
 \hat{L}_{\beta}^b \hat{E}_k^{\scriptscriptstyle C} \Big] + 
\frac{1}{2}\, \lambda_{ijk}^{\!\prime\prime}  
\hat{U}_i^{\scriptscriptstyle C} \hat{D}_j^{\scriptscriptstyle C}  
\hat{D}_k^{\scriptscriptstyle C}   ,
\eeqa\normalsize
where  $(a,b)$ are $SU(2)$ indices, $(i,j,k)$ are the usual family (flavor) 
indices, and $(\za, \zb)$ are extended flavor index going from $0$ to $3$.
In the limit where $\lambda_{ijk}, \lambda^{\!\prime}_{ijk},  
\lambda^{\!\prime\prime}_{ijk}$ and $\mu_{i}$  all vanish, 
one recovers the expression for the R-parity preserving case, 
with $\hat{L}_{0}$ identified as $\hat{H}_d$. Without R-parity imposed,
the latter is not {\it a priori} distinguishable from the $\hat{L}_{i}$'s.
Note that $\lambda$ is antisymmetric in the first two indices, as
required by  the $SU(2)$  product rules, as shown explicitly here with 
$\varepsilon_{\scriptscriptstyle 12} =-\varepsilon_{\scriptscriptstyle 21}=1$.
Similarly, $\lambda^{\!\prime\prime}$ is antisymmetric in the last two 
indices, from $SU(3)_{\scriptscriptstyle C}$. 

R-parity is exactly an {\it ad hoc} symmetry put in to make $\hat{H}_d$
stand out from the other $\hat{L}_i$'s. It is defined in terms of
baryon number, lepton number, and spin as, explicitly, 
${\mathcal R} = (-1)^{3B+L+2S}$. The consequence is that the accidental 
symmetries of baryon number and lepton number in the SM are preserved, at
the expense of making particles and superparticles having a categorically 
different quantum number, R-parity. The latter is actually 
not the most effective discrete symmetry to control superparticle 
mediated proton decay\cite{pd}, but is most restrictive for term admitted
in the superpotential.

Doing phenomenological studies without specifying a choice 
of flavor bases is ambiguous. It is like doing SM quark physics with 18
complex Yukawa couplings, instead of the 10 real physical parameters.
As far as the SM itself is concerned, the extra 26 real parameters
are simply redundant, and attempts to relate the full 36 parameters to
experimental data will be futile.
In SUSY without R-parity, the choice of an optimal
parametrization mainly concerns the 4 $\hat{L}_\alpha$ flavors. Under the SVP, 
flavor bases are chosen such that : 
1/ among the $\hat{L}_\alpha$'s, only  $\hat{L}_0$, bears a VEV,
{\it i.e.} {\small $\langle \hat{L}_i \rangle \equiv 0$};
2/  {\small $h^{e}_{jk} (\equiv \lambda_{0jk}) 
=\frac{\sqrt{2}}{v_{\scriptscriptstyle 0}} \,{\rm diag}
\{m_{\scriptscriptstyle 1},
m_{\scriptscriptstyle 2},m_{\scriptscriptstyle 3}\}$};
3/ {\small $h^{d}_{jk} (\equiv \lambda^{\!\prime}_{0jk} =-\lambda_{j0k}) 
= \frac{\sqrt{2}}\,{v_{\scriptscriptstyle 0}}{\rm diag}\{m_d,m_s,m_b\}$}; 
4/ {\small $h^{u}_{ik}=\frac{v_{\scriptscriptstyle u}}{\sqrt{2}}
V_{\!\mbox{\tiny CKM}}^{\!\scriptscriptstyle T}\;
{\rm diag}\{m_u,m_c,m_t\}$}, where 
${v_{\scriptscriptstyle 0}} \equiv  \sqrt{2}\,\langle \hat{L}_0 \rangle$
and ${v_{\scriptscriptstyle u} } \equiv \sqrt{2}\,
\langle \hat{H}_{u} \rangle$. 
The big advantage of here is that the (tree-level) mass 
matrices for {\it all} the fermions  {\it do not} involve any of
the trilinear RPV couplings, though the approach makes {\it no assumption} 
on any RPV coupling including even those from soft SUSY breaking;
and all the parameters used are uniquely defined, with the exception of
some removable phases. 
In fact, the (color-singlet) charged fermion mass 
matrix is reduced to the simple form :
\beq \label{mc}
{\mathcal{M}_{\scriptscriptstyle C}} =
 \left(
{\begin{array}{ccccc}
{M_{\scriptscriptstyle 2}} &  
\frac{g_{\scriptscriptstyle 2}{v}_{\scriptscriptstyle 0}}{\sqrt 2}  
& 0 & 0 & 0 \\
 \frac{g_{\scriptscriptstyle 2}{v}_{\scriptscriptstyle u}}{\sqrt 2} & 
 {{ \mu}_{\scriptscriptstyle 0}} & {{ \mu}_{\scriptscriptstyle 1}} &
{{ \mu}_{\scriptscriptstyle 2}}  & {{ \mu}_{\scriptscriptstyle 3}} \\
0 &  0 & {{m}_{\scriptscriptstyle 1}} & 0 & 0 \\
0 & 0 & 0 & {{m}_{\scriptscriptstyle 2}} & 0 \\
0 & 0 & 0 & 0 & {{m}_{\scriptscriptstyle 3}}
\end{array}}
\right)  \; .
\eeq
Readers are referred to Ref.\cite{k} for details conerning the RPV
effects on the leptons.
%

\noindent{\it Squark mixing and EDM.}
The soft SUSY breaking part of the Lagrangian can be written as follows :
\beqa
V_{\rm soft} &=& \epsilon_{\!\scriptscriptstyle ab} 
  B_{\za} \,  H_{u}^a \tilde{L}_\za^b +
\epsilon_{\!\scriptscriptstyle ab} \left[ \,
A^{\!\scriptscriptstyle U}_{ij} \, 
\tilde{Q}^a_i H_{u}^b \tilde{U}^{\scriptscriptstyle C}_j 
+ A^{\!\scriptscriptstyle D}_{ij} 
H_{d}^a \tilde{Q}^b_i \tilde{D}^{\scriptscriptstyle C}_j  
+ A^{\!\scriptscriptstyle E}_{ij} 
H_{d}^a \tilde{L}^b_i \tilde{E}^{\scriptscriptstyle C}_j   \,
\right] + {\rm h.c.}\nonumber \\
&+&
\epsilon_{\!\scriptscriptstyle ab} 
\left[ \,  A^{\!\scriptscriptstyle \lambda^\prime}_{ijk} 
\tilde{L}_i^a \tilde{Q}^b_j \tilde{D}^{\scriptscriptstyle C}_k  
+ \frac{1}{2}\, A^{\!\scriptscriptstyle \lambda}_{ijk} 
\tilde{L}_i^a \tilde{L}^b_j \tilde{E}^{\scriptscriptstyle C}_k  
\right] 
+ \frac{1}{2}\, A^{\!\scriptscriptstyle \lambda^{\prime\prime}}_{ijk} 
\tilde{U}^{\scriptscriptstyle C}_i  \tilde{D}^{\scriptscriptstyle C}_j  
\tilde{D}^{\scriptscriptstyle C}_k  + {\rm h.c.}
\nonumber \\
&+&
 \tilde{Q}^\dagger \tilde{m}_{\!\scriptscriptstyle {Q}}^2 \,\tilde{Q} 
+\tilde{U}^{\dagger} 
\tilde{m}_{\!\scriptscriptstyle {U}}^2 \, \tilde{U} 
+\tilde{D}^{\dagger} \tilde{m}_{\!\scriptscriptstyle {D}}^2 
\, \tilde{D} 
+ \tilde{L}^\dagger \tilde{m}_{\!\scriptscriptstyle {L}}^2  \tilde{L}  
  +\tilde{E}^{\dagger} \tilde{m}_{\!\scriptscriptstyle {E}}^2 
\, \tilde{E}
+ \tilde{m}_{\!\scriptscriptstyle H_{\!\scriptscriptstyle u}}^2 \,
|H_{u}|^2 
\nonumber \\
&& + \frac{M_{\!\scriptscriptstyle 1}}{2} \tilde{B}\tilde{B}
   + \frac{M_{\!\scriptscriptstyle 2}}{2} \tilde{W}\tilde{W}
   + \frac{M_{\!\scriptscriptstyle 3}}{2} \tilde{g}\tilde{g}
+ {\rm h.c.}\; ,
\label{soft}
\eeqa
where we have separated the R-parity conserving ones from the 
RPV ones ($H_{d} \equiv \hat{L}_0$) for the $A$-terms. Note that 
$\tilde{L}^\dagger \tilde{m}_{\!\scriptscriptstyle \tilde{L}}^2  \tilde{L}$,
unlike the other soft mass terms, is given by a 
$4\times 4$ matrix. Explicitly, 
$\tilde{m}_{\!\scriptscriptstyle {L}_{00}}^2$ is
$\tilde{m}_{\!\scriptscriptstyle H_{\!\scriptscriptstyle d}}^2$ 
of the MSSM case while 
$\tilde{m}_{\!\scriptscriptstyle {L}_{0k}}^2$'s give RPV mass mixings.

We have illustrated above how the SVP keeps the expressions for the down-quark 
and color-singlet charged fermion mass matrices simple. The SVP performs 
the same trick to the corresponding scalar sectors as
well. Here, we concentrate on the down-squarks. We have the 
mass-squared matrix as follows : 
\beq
{\cal M}_{\!\scriptscriptstyle {D}}^2 =
\left( \begin{array}{cc}
{\cal M}_{\!\scriptscriptstyle LL}^2 & {\cal M}_{\!\scriptscriptstyle RL}^{2\dag} \\
{\cal M}_{\!\scriptscriptstyle RL}^{2} & {\cal M}_{\!\scriptscriptstyle RR}^2
 \end{array} \right) \; ,
\eeq
where ${\cal M}_{\!\scriptscriptstyle LL}^2$ and  
${\cal M}_{\!\scriptscriptstyle RR}^2$ are the same as in MSSM 
while
\beq \label{RL}
({\cal M}_{\!\scriptscriptstyle RL}^{2})^{\scriptscriptstyle T} = 
A^{\!{\scriptscriptstyle D}} \frac{v_{\scriptscriptstyle 0}}{\sqrt{2}}
- m_{\!\scriptscriptstyle D} \, \mu_{\scriptscriptstyle 0}^* \, \tan\!\beta 
- (\, \mu_i^*\lambda^{\!\prime}_{ijk}\, )_\star \; 
\frac{v_{\scriptscriptstyle u}}{\sqrt{2}} \; .
\eeq
Here, $m_{\!\scriptscriptstyle D}$ is the down-quark mass matrix, 
which is diagonal under the parametrization adopted; 
$(\, \mu_i^*\lambda^{\!\prime}_{ijk}\, )_\star$ denotes the 
$3\times 3$ matrix $(\;)_{jk}$ with elements listed; and 
$\tan\!\beta =\frac{v_{\scriptscriptstyle u}}{v_{\scriptscriptstyle 0}}$.
Note that in the equation for
$({\cal M}_{\!\scriptscriptstyle RL}^2)^{\scriptscriptstyle T}$, 
we can write the first, $A$-term, contribution as
\beq  \label{A}
A^{\!{\scriptscriptstyle D}}\, \frac{v_{\scriptscriptstyle 0}}{\sqrt{2}}
= A_d \,m_{\!\scriptscriptstyle D} +\delta\! A^{\!{\scriptscriptstyle D}} 
\, \frac{v_{\scriptscriptstyle 0}}{\sqrt{2}}
\eeq
with $A_d$ being a constant (mass) parameter representing the 
``proportional" part. The remaining terms in 
$({\cal M}_{\!\scriptscriptstyle RL}^2)^{\scriptscriptstyle T}$ 
are $F$-term contributions; in particular, the last term 
is a ``SUSY conserving"\cite{foot} 
but R-parity violating contributions given here for the first time. 
In fact, contributions to $LR$ scalar mixing of this type,
for the sleptons, is first identified in a recent paper\cite{ck} 
where their role in  the SUSY analog of the  Zee neutrino mass 
model\cite{zee} is discussed. In a parallel paper by one of the authors 
(O.K.)\cite{kong}, a systematic analysis of the full squark and slepton
masses as well as their contributions, through $LR$ mixings,  to 1-loop 
neutrino masses are also presented. Here, we focus only on the down-quark 
sector. Note that the full $F$-term part in the above equation can
actually be written together as 
$(\, \mu_{\alpha}^*\lambda^{\!\prime}_{{\alpha}jk}\, )_\star \; 
\frac{v_{\scriptscriptstyle u}}{\sqrt{2}}$ where the $\alpha=0$ 
term gives the second term in RHS of Eq.(\ref{RL}), which 
is the usual $\mu$-term contribution in the MSSM case. The 
latter is, however, diagonal, {\it i.e.} vanishes for $j \ne k$.
We would like to emphasize that the above 
result is complete --- all RPV contributions are included. The 
simplicity of the result is a consequence of the SVP.
Explicitly, the RPV $A$-terms contributions [{\it c.f.} Eq.(\ref{A})]
vanish as $v_i\equiv \sqrt{2}\langle\hat{L}_i\rangle=0$

The $(\, \mu_i^*\lambda^{\!\prime}_{ijk}\, )_\star$ term is very interesting.
It involves only parameters in the superpotential and has  {\it nothing to do 
with soft SUSY breaking}. Without an underlining flavor theory, there is
no reason to expect any specific structure among different terms of the
matrix. In particular, the off-diagonal terms ($j\ne k$) may have an important
role to play. They contribute to flavor changing neutral current 
(FCNC) processes such as $b\to s\,\gamma$, a topic to be addressed in a later
publication\cite{next}. Moreover, both the $\mu_i$'s and the 
$\lambda^{\!\prime}_{ijk}$'s are complex parameters. Hence, diagonal terms in 
$(\, \mu_i^*\lambda^{\!\prime}_{ijk}\, )_\star$ also bear 
CP-violating phases and contribute to electric dipole moments (EDM's) of 
the corresponding quarks. In particular,  
$\mu_i^*\lambda^{\!\prime}_{i\scriptscriptstyle 1\!1}$ gives contribution
to neutron EDM at 1-loop level, in exactly the same fashion as 
the $A$-term in MSSM does. The similar term in $LR$ 
slepton mixing gives rise to electron EDM. This result is in direct 
contrary to the impression one may get from reading the two recent papers
on the subject\cite{two}. One should bear in mind that the two papers 
do not put together both the bilinear and the trilinear RPV terms. 
Our treatment here, bases on the SVP, gives, for the first time, 
the result of squark masses for the complete theory of SUSY without R-parity. 
Going from here, obtaining the EDM contributions is straight forward.

Contribution to EDM of the $d$ quark at 1-loop level, from a gaugino loop with 
$LR$-squark mixing in particular (see Fig.~1), has been widely studied within 
MSSM\cite{susycp,KiOs,INa,BS}. 
With the squark mixings in the down-sector parametrized by 
$\delta_{jk}^{\scriptscriptstyle D}$ (normalized by average squark mass as
explicitly shown below), we have the neutron EDM result given by
\beq
d_{\scriptscriptstyle n} = -\frac{8}{27} \frac{e\, \alpha_s}{\pi}\,
\frac{M_{\scriptscriptstyle \tilde{g}}}{M_{\scriptscriptstyle \tilde{d}}^2}\,
\mbox{Im}(\delta_{\scriptscriptstyle 1\!1}^{\scriptscriptstyle D})\;
F_1\!\left(\frac{M_{\scriptscriptstyle \tilde{g}}^2}
{M_{\scriptscriptstyle \tilde{d}}^2}\right) \;
\eeq
where $M_{\scriptscriptstyle \tilde{g}}$ and
$M_{\scriptscriptstyle \tilde{d}}$ are the gluino and down squark masses
respectively, and
\beq
F_1(x)= \frac{1}{(1-x)^3} \left( \frac{1+5x}{2} + 
\frac{2+x}{1-x} \ln{x} \right) \; .
\eeq 
Contribution of $\mu_i^*\lambda^{\!\prime}_{i\scriptscriptstyle 1\!1}$
to $\delta_{\scriptscriptstyle 1\!1}^{\scriptscriptstyle D}$
is to be given as
\[
- \mu_i^*\lambda^{\!\prime}_{i\scriptscriptstyle 1\!1}\,
\frac{v_{\scriptscriptstyle u}}{\sqrt{2}}\,
\frac{1}{M_{\scriptscriptstyle \tilde{d}}^2}\;.
\]
Requiring the contribution alone not to upset the experimental bound on 
neutron EDM : 
$(d_n)^{\mbox{\tiny exp}} < 6.3 \cdot 10^{-26}\,e \cdot \mbox{cm}$, 
a bound can be obtained for the RPV parameters.
Note that going from $d$ quark EDM to neutron EDM, we assume the simple 
valence quark model\cite{HKP}.
Taking $M_{\scriptscriptstyle \tilde{d}}=100\,\mbox{GeV}$ 
and $M_{\scriptscriptstyle \tilde{g}}=300\,\mbox{GeV}$
gives the bound
\beq \label{bound}
\mbox{Im}(\mu_i^*\lambda^{\!\prime}_{i\scriptscriptstyle 1\!1}) 
\leq 10^{-6}\,\mbox{GeV} \; ,
\eeq
(with $v_{u} \sim 200\,\mbox{GeV}$).
This result is interesting. Let us first concentrate on the $i=3$ part, 
assuming the $i=1$ and $2$ contribution to be subdominating. 
Imposing the $18.2\,\mbox{MeV}$ experimental 
bound\cite{aleph} for the mass of $\nu_\tau$ still admits a relatively
large $\mu_{\scriptscriptstyle 3}$, especially for a large $\tan\!\zb$.
Reading from the results in Ref.\cite{k}, the bound is $\sim 7\,\mbox{GeV}$ 
at $\tan\!\zb=2$ and $\sim 300\,\mbox{GeV}$ at $\tan\!\zb=45$, 
while the best bound on the corresponding 
$\lambda^{\!\prime}_{\scriptscriptstyle 31\!1}$ (from
$\tau \to \pi \nu$) is around $0.05\sim 0.1$\cite{lambda}. 

Here, an explicit comparison with the corresponding R-parity conserving 
contribution is of interest. From Eqs.(\ref{RL}) and (\ref{A}), it is obvious 
that we are talking about 
$(A_d - \mu_{\scriptscriptstyle 0}^* \, \tan\!\beta) \, m_{\scriptscriptstyle d}$ 
verses $-  \mu_i^*\lambda^{\!\prime}_{i\scriptscriptstyle 1\!1}\,  
\frac{v_{\scriptscriptstyle u}}{\sqrt{2}}$. Both $A_d$ and 
$\mu_{\scriptscriptstyle 0}$ are expected to be roughly at the same order as
$v_{\scriptscriptstyle u}$, {\it i.e.} at electroweak scale. We are hence
left to compare $m_{\scriptscriptstyle d}$ ($\sim 10^{-3}\,\mbox{GeV}$)
with $\mu_i^*\lambda^{\!\prime}_{i\scriptscriptstyle 1\!1}$. The above
discussion then leads to the conclusion that the RPV part could easily
be larger by one or even two orders of magnitude. 

On the other hand, if one insists on a sub-eV mass for $\nu_\tau$ 
as suggested, but far from mandated, by the result from the 
Super-Kamiokande (super-K) experiment\cite{sK}, we would have 
$\mu_{\scriptscriptstyle 3} \cos\!\zb \leq 10^{-4}\,\mbox{GeV}$\cite{knu}.
This means that at least for the large $\tan\!\zb$ case, the EDM bound 
as given by Eq.(\ref{bound}) still worths notification, even under this
most limiting scenario. 

The $\mu_i^*\lambda^{\!\prime}_{i\scriptscriptstyle 1\!1}$ contribution
to squark mixing, as well as $\lambda^{\!\prime}_{i\scriptscriptstyle 1\!1}$
in itself together with an $A$-term mixing, also gives rise to
neutrino mass at 1-loop. Hence, to consistently impose the 
super-K sub-eV neutrino mass scenario, one should also check
the corresponding bound obtained. We are interested here in whether 
these will further weaken the implication of the EDM bound discussed
here. Fig.~2 shows a familiar quark-squark loop neutrino
mass diagram. We are interested here in the case where both the 
$\lambda^{\!\prime}$-couplings are
$\lambda^{\!\prime}_{\scriptscriptstyle 31\!1}$. We have, for the
R-parity conserving $LR$ squark mixing, the familiar result
\beq \label{nuA}
 m_{\nu_\tau} \sim 
\frac{3}{8\pi^2}\, m_{\scriptscriptstyle d}^2 \,
\frac{ ( A_d - \mu_{\scriptscriptstyle 0}^* \tan\!\beta)}
{M_{\scriptscriptstyle \tilde{d}}^2}\;
\lambda^{\!\prime}_{\scriptscriptstyle 31\!1} \; .
\eeq
However, with the full $LR$ mixing result as given in Eq.(\ref{RL}),
there is an extra contribution to be given as
\beq \label{nunew}
\frac{3}{8\pi^2}\, m_{\scriptscriptstyle d} \,
\frac{v_{\scriptscriptstyle u}}{\sqrt{2}}\,
\frac{\mu_i^*\lambda^{\!\prime}_{i\scriptscriptstyle 1\!1}}
{M_{\scriptscriptstyle \tilde{d}}^2}\;
\lambda^{\!\prime\,2}_{\scriptscriptstyle 31\!1} \; .
\eeq
The latter type of RPV contribution to neutrino masses has not
been identified before (see however Refs\cite{ck} and \cite{kong}).

From Eq.(\ref{nuA}), one can easily see that the requirement for
$m_{\nu_\tau}$ to be at the super-K atmospheric neutrino oscillation
scale only gives a bound for 
$\lambda^{\!\prime}_{\scriptscriptstyle 31\!1}$ of about the same
magnitude as one quoted above, from the other sources. As for the
contribution [Eq.(\ref{nunew})], the bound given by 
Eq.(\ref{bound}) itself says the contribution is smaller than
the previous one. Hence, neutrino mass contributions from
Fig.~2 do not change our conclusion above. 

Note that the EDM bound given by Eq.(\ref{bound}) actually involves 
a summation over index $i$. Results from Ref.\cite{k} indicated
that while $\mu_{\scriptscriptstyle 1}$ is very strongly bounded,
the bound on $\mu_{\scriptscriptstyle 2}$ could be not very strong. 
Moreover, the bound on $\lambda^{\!\prime}_{\scriptscriptstyle 21\!1}$
is no better than that on 
$\lambda^{\!\prime\,2}_{\scriptscriptstyle 31\!1}$\cite{lambda}.
Hence, the EDM bound may still be of interest there too. 
The story for imposing the super-K constraint is obviously the same
as the above discussion for the $i=3$ case.

One should bear in mind that the
EDM and the neutrino mass bounds 
involve different combinations of RPV parameters as well as with the other 
SUSY parameters.
An exact comparison for bounds obtained from the two sources is hence 
difficult. Our discussion above is aimed at illustrating the fact that the 
EDM bound is not completely overshadowed by the super-K neutrino mass bound.
In other word, even requiring the magnitudes for the RPV parameters to
satisfy the most stringently interpreted super-K bounds does not make
them so small that the above discussed contribution to neutron EDM will
always be satisfied. 

\noindent{\it Beyond the gluino diagram.}
Similar RPV contributions on the neutron and electron EDM's
are obtained through neutralino exchange diagram. One simply has to replace
the gluino in the diagram with the other neutral gauginos. In the neutron case,
the gluino diagram contribution discussed here no doubt dominate,
due to the much stronger QCD coupling. 

There are other 1-loop contributions. In the case of MSSM, the chargino
contribution is known to be competitive or even dominates over the gluino
one in some regions of the parameter space\cite{KiOs}. The major part of the
chargino contribution comes from a diagram with a gauge and a Yukawa coupling
for the loop vertices, with pure $L$-squark running in the loop. Here we
give the corresponding formula generalized to the case of SUSY without 
R-parity. This is given by\cite{as6}
\begin{equation} \label{edmco}
\left({d_{\scriptscriptstyle f} \over e} \right)_{\!\!\chi^{\!\!\mbox{ -}}} = -
{\alpha_{\mbox{\tiny em}} \over 4 \pi \,\sin\!^2\theta_{\!\scriptscriptstyle W}} \; 
\sum_{\scriptscriptstyle \tilde{f}'\mp} 
\sum_{n=1}^{5} \,\mbox{Im}({\cal C}_{\!fn\mp}) \;
{{M}_{\!\scriptscriptstyle \chi^{\mbox{-}}_n} \over 
M_{\!\scriptscriptstyle \tilde{f}'\mp}^2} \;
\left[ {\cal Q}_{\!\tilde{f}'} \; 
B\!\left({{M}_{\!\scriptscriptstyle \chi^{\mbox{-}}_{n}}^2 \over 
M_{\!\scriptscriptstyle \tilde{f}'\mp}^2} \right) 
+ ( {\cal Q}_{\!{f}} - {\cal Q}_{\!\tilde{f}'} ) \;
A\!\left({{M}_{\!\scriptscriptstyle \chi^{\mbox{-}}_{n}}^2 \over 
M_{\!\scriptscriptstyle \tilde{f}'\mp}^2} \right) 
\right] \; ,
\end{equation}
for $f$ being $u$ ($d$) quark and $f'$ being $d$ ($u$), where
\beqa
{\cal C}_{un\mp} &=&  
{y_{\!\scriptscriptstyle u} \over g_{\scriptscriptstyle 2} } \,\, 
\mbox{\boldmath $V$}^{\!*}_{\!\!2n} \, {\cal D}_{d1\mp} \;
\left(  \mbox{\boldmath $U$}_{\!1n} \,{\cal D}^{*}_{d1\mp} -
{y_{\!\scriptscriptstyle d} \over g_{\scriptscriptstyle 2} }\,\, 
\mbox{\boldmath $U$}_{\!2n}\,  {\cal D}^{*}_{d2\mp}
- {\lambda^{\!\prime}_{\scriptscriptstyle i11} \over g_{\scriptscriptstyle 2} }\,\, 
\mbox{\boldmath $U$}_{\!(i+2)n}\,  {\cal D}^{*}_{d2\mp} \right) \; ,
\nonumber \\
{\cal C}_{dn\mp} &=& 
\left( {y_{\!\scriptscriptstyle d} \over g_{\scriptscriptstyle 2} }\,\, 
\mbox{\boldmath $U$}_{\!2n} 
+ {\lambda^{\!\prime}_{\scriptscriptstyle i11} \over g_{\scriptscriptstyle 2} }\,\, 
\mbox{\boldmath $U$}_{\!(i+2)n} \right)\! {\cal D}_{u1\mp} \;
\left( \mbox{\boldmath $V$}^{\!*}_{\!\!1n} \,{\cal D}^{*}_{u1\mp} -
{y_{\!\scriptscriptstyle u} \over g_{\scriptscriptstyle 2} } \,
\mbox{\boldmath $V$}^{\!*}_{\!\!2n} \, {\cal D}^{*}_{u2\mp} \right) 
\; .
\label{Cnmp}
\eeqa
The terms in ${\cal C}_{dn\mp}$ with only one factor of 
${1\over g_{\scriptscriptstyle 2} }$ and a 
$\lambda^{\!\prime}_{i\scriptscriptstyle 1\!1}$ gives the RPV analog of
the dominating MSSM chargino contribution. The term
is described by a diagram, which at first order requires a 
${l}_{\scriptscriptstyle L_i}^{\!\!\mbox{ -}}$--$\tilde{W}^{\scriptscriptstyle +}$ 
mass mixing. The latter vanishes, as shown in Eq.(\ref{mc}). From the full
formula above, it is easy to see that the exact mass
eigenstate result would deviate from zero only to the extent that the mass
dependence of the $B$ and $A$ functions\cite{as6} spoils the GIM like
cancellation in the sum. The resultant contribution, however, is shown by
our exact numerical calculation to be substantial.
What is most interesting here is that an analysis 
through perturbational approximations illustrates that the contribution
is proportional to, basically, the same combination of RPV parameters,
{\it i.e.} $\mu_i^* \, \lambda^{\!\prime}_{\scriptscriptstyle i11}$.
While we cannot give much of the details here (see Ref.\cite{as6}), let us
list numbers from a sample point for illustration : with $A_u=A_d=500\,\mbox{GeV}$,
${\mu}_{\scriptscriptstyle 0}=-300\,\mbox{GeV}$, $\tan\!\zb = 3$, a
common gaugino masses at $300\,\mbox{GeV}$, 
$\tilde{m}_{\scriptscriptstyle Q}=200\,\mbox{GeV}$,
$\tilde{m}_{\scriptscriptstyle u}=
 \tilde{m}_{\scriptscriptstyle d}=100\,\mbox{GeV}$, 
${\mu}_{\scriptscriptstyle 3}=1\times 10^{-4}\,\mbox{GeV}$,
and $\lambda^{\!\prime}_{\scriptscriptstyle 31\!1}=
 0.1\times \mbox{exp}{(i{\pi/6})}$ (being the only complex parameter), 
we have the resulted neutron EDM contributions from gluino, chargino(-like),
and neutralino(-like) 1-loop diagrams given by
$2.49$, $0.56$, and $-0.056$ times $10^{-27}\,e\,\mbox{cm}$, respectively.
 
\noindent{\it Summary.}
In summary, we have presented the complete result for LR squark mixing 
and analyzed its contribution to neutron EDM through the gluino diagram.
The result provide interesting new bounds on RPV parameters.
A brief discussion for the chargino(-like) 1-loop contribution is also
given, together with a sample result from exact numerical calculations,
including also the neutralino(-like) loop. The full details will be report
in an publication.
 
We would also like to mention 
that there is the analogous case for the slepton mixing and electron EDM.
The latter contributions, while having a similar structure, has potential 
complications from mixings with charged Higgs. The issue is under investigation.

\noindent{\it Acknowledgment :}
Y.Y.K. wishes to thank M. Kobayashi and H.Y.Cheng for their hositality.
His work was in part supported by the National Science Council of R.O.C.
under the Grant No. NSC-89-2811-M-001-0053. O.K. wants to thank D. Chang
for discussions.

\bigskip
\bigskip

\noindent
{\bf Figure captions :}\\[.2in]
Fig. 1 --- EDM for $d$ quark at 1-loop.\\[.2in]
Fig. 2 --- Neutrino mass at 1-loop.


\begin{figure}
\vspace*{.5in}
\includegraphics{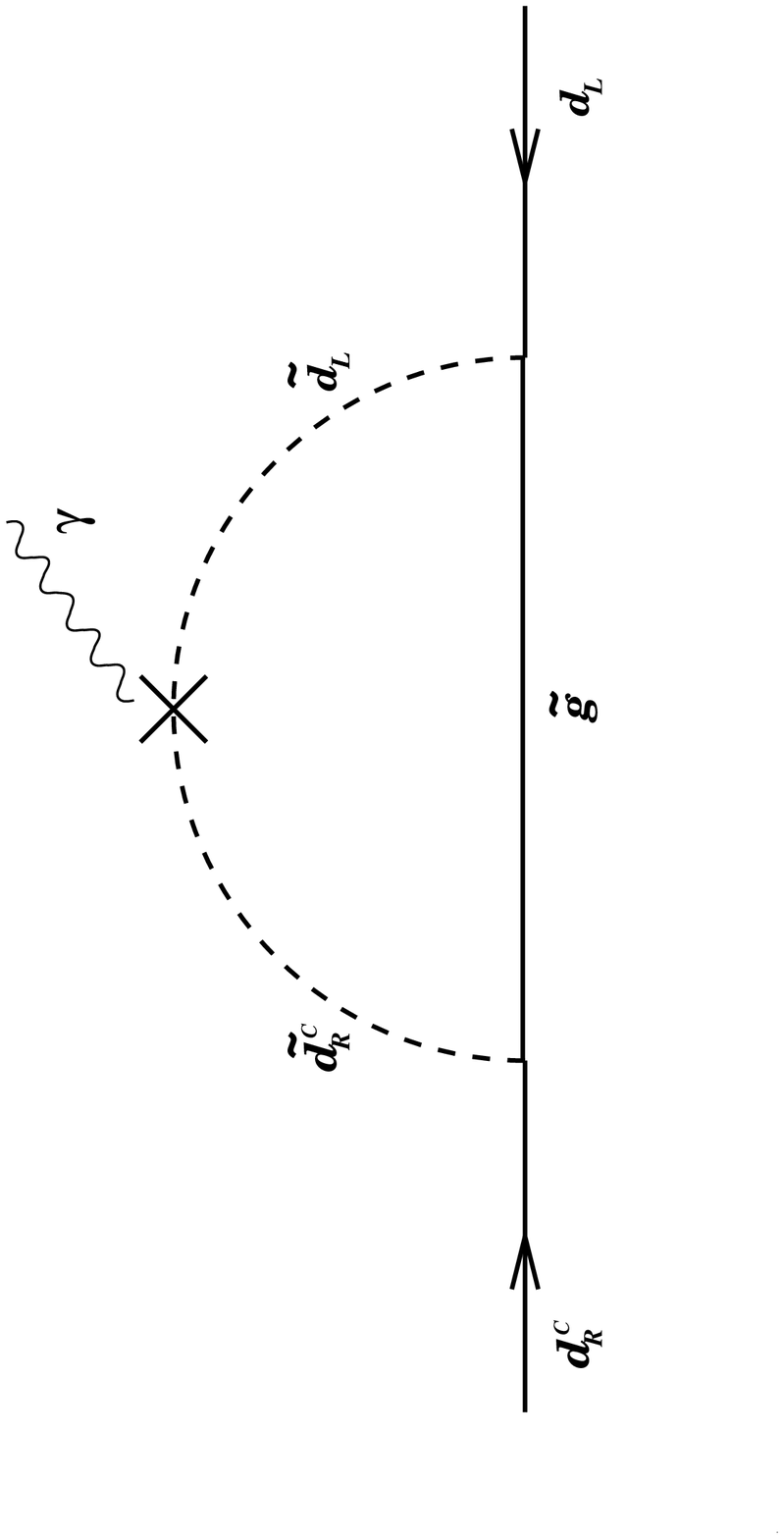}
\vspace*{3in}
\caption{EDM for $d$ quark at 1-loop.}
\end{figure}

\vspace*{1.in}

\begin{figure}
\includegraphics{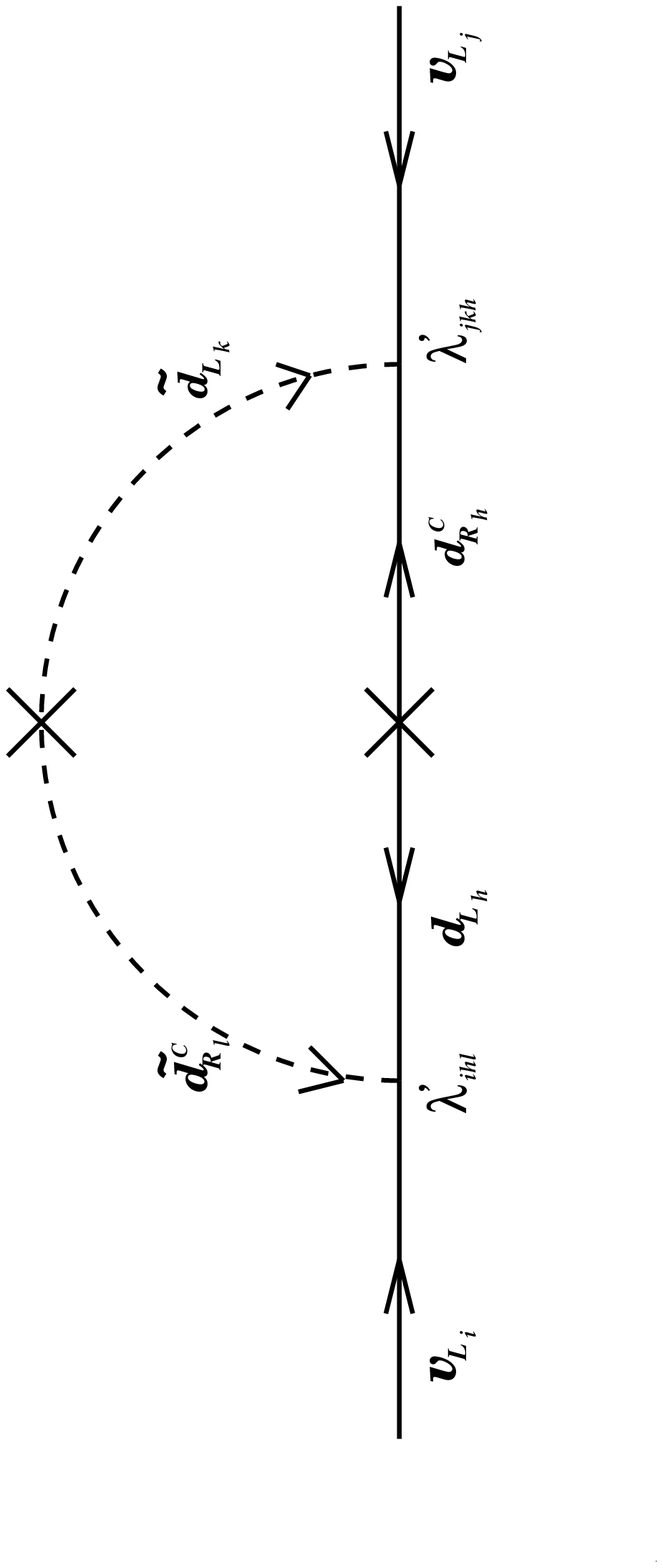}
\vspace*{3in}
\caption{Neutrino mass at 1-loop.}
\end{figure}


\begin{thebibliography}{99}
\bibitem{k}
M. Bisset, O.C.W. Kong, C. Macesanu, and L.H. Orr,
Phys. Lett. {\bf B430}, 274 (1998); 
Phys. Rev. {\bf D62},  {\it 035001} (2000).
\bibitem{susycp}
See, for example, T. Falk and K.A. Olive,
Phys. Lett. {\bf B439}, 71 (1998), and references therein.
\bibitem{two}
R.M. Godbole, S. Pakvasa, S.D. Rindani, and X. Tata,
Phys. Rev. {\bf D61},  {\it 113003} (2000);
S.A. Abel, A.Dedes, and H.K. Dreiner,  JHEP {\bf 05}, 013 (2000).
\bibitem{pd}
L.E. Ib\'a\~nez and G.G. Ross,
Nucl. Phys. {\bf B368}, 3 (1992).
\bibitem{foot}
However, it should be noted that existence of nonzero
$F$-terms or electroweak symmetry breaking VEV's can be interpreted 
as a consequence of SUSY breaking.
\bibitem{ck}
K. Cheung and O.C.W. Kong, Phys. Rev.  {\bf D61},  {\it 113012} (2000).
\bibitem{zee}
A. Zee, Phys. Lett. {\bf 93B}, 389 (1980).
\bibitem{kong}
O.C.W. Kong,  JHEP {\bf 09}, 037 (2000).
\bibitem{next}
O.C.W. Kong {\it et al.}, {\it work in progress}.
\bibitem{KiOs}
Y. Kizukuri and N. Oshimo, Phys. Rev. {\bf D46}, 
3025 (1992).
\bibitem{INa} 
T. Ibrahim and P. Nath, Phys. Rev. {\bf D57} 478 (1998),
{\it Errata : --- ibid} {\bf D58} 019901 (1998),
{\bf D60} 079903 (1999),
{\bf D60} 119901 (1999);
{\it ibid} {\bf D58} 111301 (1998);
{\it Erratum --- ibid} {\bf D60} 099902 (1999);
T. Goto, Y.-Y. Keum, T. Nihei, Y. Okada and Y. Shimizu,
Phys. Lett. {\bf 460B}, 333 (1999).
\bibitem{BS}
For a nice simple summary, see I.I. Bigi and A.I. Sanda,
{\it CP Violation}, Cambridge University Press (2000).
\bibitem{HKP}
See, for a review of the issue, 
X.-G. He, B.H.J. McKellar, and S. Pakvasa,
Int. J. Mod. Phys. {\bf A4}, 5011 (1989).
\bibitem{aleph}
R. Barate {\it et al}. (ALEPH Collaboration),
CERN-PPE-97-138, (1997).
\bibitem{lambda}
See, for example,
G. Bhattacharyya, Nucl. Phys. Proc. Suppl. {\bf 52A}, 83 (1997);
 V. Bednyakov, A. Faessler, and S. Kovalenko,  hep-ph/9904414.
\bibitem{sK}
Super-Kamiokande Collaboration, Y.Fukuda {\it et al.},
 Phys. Rev. Lett. {\bf 81} (1998) 1562;
 P. Lipari, hep-ph/9904443;
G.L. Fogli, E. Lisi, A. Marrone, and G. Scioscia,
Phys. Rev. {\bf D59} (1999) 033001.
\bibitem{knu}
O.C.W. Kong, Mod. Phys. Lett. A. {\bf 14}, 903 (1999).
\bibitem{as6} 
Details are given in Y.-Y. Keum and O.C.W. Kong, {IPAS-HEP-k006},
{\it manuscript in preparation}, with full results from numerical
calculations. Eq.(\ref{edmco}) may also be
matched with the MSSM formula in Ref.\cite{KiOs} for better understanding.
\end{thebibliography}
\end{document}